\documentclass[prl,twocolumn,showpacs]{revtex4}
\usepackage{amsmath,bbm,epsfig}
\usepackage{amsfonts}
\usepackage{mathrsfs}
\newcommand{\ii}{\;\! \mbox{i} \;\!}

\newcommand{\mb}[1]{{\rm #1}}

\begin{document} 

\title{Resonant nonlinear 
quantum transport for a periodically kicked Bose condensate}

\author{Sandro Wimberger, Riccardo Mannella, Oliver Morsch, 
and Ennio Arimondo}  
\affiliation{INFM, Dipartimento di Fisica E. Fermi,
Unversit\`a di Pisa, Largo Pontecorvo 3, 56127 Pisa, Italy}

\begin{abstract}
Our realistic numerical results show that the fundamental and 
higher-order quantum resonances of the $\delta$-kicked rotor are observable
in state-of-the-art experiments with a Bose condensate in a shallow harmonic
trap, kicked by a spatially periodic optical lattice. For stronger
confinement, interaction-induced destruction of the resonant motion 
of the kicked harmonic oscillator is predicted.
\end{abstract}


\maketitle

Dynamical systems that exchange significant energy with an 
external driving (time-dependent) field are paradigmatic objects for the
study of complex evolutions with only a few degrees of freedom. In contrast
to autonomous systems, where chaotic behavior can originate from 
many-body interactions, the complexity in driven systems arises from 
and can be controlled by the external field. 
An experimental setup where {\em both}
types of complexity -- many-body dynamics and external drive -- are present
is realizable and to a large degree controllable
with state-of-the-art atom optical systems 
\cite{Phillips1999,Wilson2004AR}. 
Good control over nontrivial dynamics is the necessary
tool for manipulating quantum states in a desired way \cite{CC}. 
Controlled coherent evolution is the necessary ingredient for quantum
computing schemes, and, in particular, for algorithms which are based on 
a fast spreading over the Hilbert space of interest. Recently, ballistic
expansion (i.e., linearly increasing momenta with time) was proposed
to realize such quantum random walk algorithms \cite{QRW}. 

In this Letter, we answer the question of whether ballistic resonant
quantum transport can be realized with a periodically kicked Bose condensate.
Cold dilute atomic gases have so far been used to realize many features
of quantum chaos, such as dynamical localization \cite{expold}
or dynamical tunneling \cite{DT}.
To implement a fast spreading in momentum space, one can use quantum
accelerator modes found recently \cite{AC}, 
or the standard $\delta$-kicked rotor (KR) which shows 
ballistic motion at the so-called quantum resonances
\cite{Izr}.
The latter, however, are hard to realize if the initial conditions cannot be
optimally controlled \cite{QRexp}. 
The preparation of a Bose condensate within harmonic
traps offers very well-defined initial momenta, necessary for observing 
ballistic motion for a substantial number of kicks.

We use the time-dependent Gross-Pitaevskii equation (GPE) \cite{PS2002} to
describe a Bose-Einstein condensate in an harmonic 
confinement which is subject to a temporally and spatially periodic optical
potential, created by a far detuned optical lattice. 
If the external potentials are not too strong, the
GPE provides a good approximation of experiments
with dilute Bose gases \cite{Zoller2000,ZLRN2004}. 
The GPE which we numerically integrate has the
following form:
\begin{eqnarray}
&\ii \hbar \frac{\partial \psi (\vec{r},t)}{\partial t} =
\left[-\frac{\hbar^2 \nabla ^2}{2M} + \frac{M \omega_x^2 x^2}{2} +
\frac{M \omega_{\rm \tiny r}^2 \rho^2}{2} + V_0\cos(2k_L x) \right. 
\nonumber \\
& \left. \times \sum_{m=-\infty}^{+\infty} F(t-mT) + g N |\psi(\vec{r},t)|^2
\right] \psi(\vec{r},t)\;,
\label{eq:GP}
\end{eqnarray}
with $\rho^2 = y^2 + z^2$. $\psi(\vec{r},t)$ represents the condensate
wave function, and $M$ is the atomic mass. The nonlinear
coupling constant is given by $g=4\pi \hbar^2 a/M$, $N$ is the
number of atoms in the condensate, $a$ the s-wave scattering length, and
$k_L$ is the wave vector of the laser
creating the optical potential. In principle, an arbitrary pulse shape
$F(t)$ may be realized, but here we restrict ourselves to a situation
where the laser is switched on at time instants separated by $T$, 
with maximum amplitude $V_0$ and periodic pulse shape function $F(t)$
of unit amplitude and duration $\delta T \ll T$. Our system is then 
the nonlinear analogue of noninteracting cold atom experiments 
\cite{expold,QRexp}, which implement the KR model, and
we can directly compare our results with the well-studied 
KR (for $\omega_x \to 0$), or with the Kicked Harmonic Oscillator (KHO)
\cite{QKHO} (for non-vanishing $\omega_x$).
The commonly used dimensionless kick strength and kick period
of the KR are $k = (V_0/\hbar) \int_0^T dt\ F(t)$ and 
$\tau = 8TE_{\mb{R}}/\hbar$, 
with the recoil energy $E_{\mb{R}}=(\hbar k_L)^2/2M$ \cite{QRexp}. 

The GPE was numerically integrated using a finite difference 
propagator, adapted by a predictor-corrector loop to reliably 
evaluate the nonlinear interaction \cite{Mannella1998}. The external,
time-dependent potential makes the integration challenging, 
in both time and computer memory. Typical integration times range from
a few hours for a simplified 1D version of Eq.~(1), to several weeks
for the full 3D problem. For the 1D model,
the motion is confined to the longitudinal ($x$) direction, and we
use the renormalized  
nonlinear coupling parameter $g_{1D}=2\hbar\omega_{\mbox{\tiny r}} a$, 
assuming a radial trapping frequency $\omega_{\mbox{\tiny r}} 
\gg \omega_x$ \cite{O1998}.  
Experimentally, such a confinement 
is obtained using a cigar-shaped optical tube,
as realized in the experiment of Moritz {\em et al.} \cite{Ess2003}.
The initial state inserted into (1) is the relaxed 
condensate wave function corresponding to the 
{\em experimental ground state} in a magnetic or optical trap. 
The ground state lies between the cases of a Gaussian, for $g=0$, and
the Thomas-Fermi limit, which is essentially an inverted parabola,
for large nonlinearity \cite{PS2002}. Its characteristic width $\sigma_{p_x}$ 
in momentum space is determined by the nonlinearity. 
Increasing the nonlinearity in Eq.~(1)
leads to a smaller width $\sigma_{p_x}$ of the initial state \cite{PS2002}. 

The dynamics of the system described by (1) depends sensitively on the 
relative strength of the three potentials, i.e., on the control parameters
$V_0$ (for fixed pulse shape), $\omega_x$, and $g$. Since the system 
absorbs energy from the optical lattice, we must at all times compare the 
kinetic energy with the (longitudinal) trap potential and the nonlinear
term. If the latter two contributions are small, we can easily realize
ballistic quantum motion, up to interaction times above which 
the trap potential is no longer negligible. On the other hand, we can tune
the system to a situation where the trap crucially affects the
dynamics, and we then have a realization of the KHO. 

In the following, we show that the fundamental as well as higher-order quantum
resonances (QR) of the KR can be observed in an experiment using a Bose
condensate, in the presence of a shallow harmonic confinement. 
In the linear KR, the QRs occur at specific kick periods 
$T=T_T s/r$ ($s,r$ integer) \cite{Izr}. At the Talbot time $T=T_T$,
the amplitudes of the wave function in momentum space exactly rephase
between successive kicks for particular initial momenta 
($p_{\mbox{\tiny init}}=0$) \cite{Izr,Phillips1999}. The result is a maximal, 
i.e., perfectly phase-matched, absorption of energy from the kicks, leading to
a ballistic spread of the wave packet \cite{Izr,QRexp}.
Only signatures of the QRs at $T=T_T/2$ and $T=T_T$ have been observed
up to now in experiments with essentially non-interacting atoms, because
the initial momenta of the atoms could not be sufficiently controlled
\cite{QRexp}. 

\begin{figure}
\centerline{\epsfig{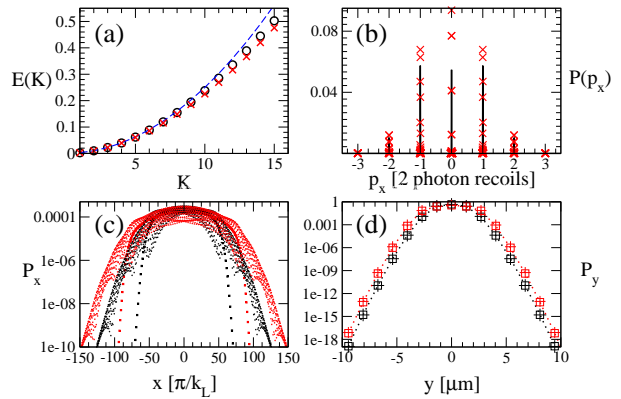}} 
\caption{(color online). (a) Kinetic energy (in units of 
$8E_{\mb{R}}$) vs $K$,
at $T=T_T$, $k=0.1$ 
(pulse width $\delta T \simeq 500 \rm \;ns$, rise time
$70\rm \;ns$; $V_0/E_{\mb{R}}\simeq 8$), 
$\omega_x/2\pi = 10 \rm \;Hz$, $\omega_{\mbox{\tiny r}} /2\pi 
=100\rm \;Hz$, and  $N=10^{4}$ (circles), $N=5\times 10^{4}$ (crosses). 
The linear KR evolution with $p_{\mbox{\tiny init}}=0$ 
is shown by the dashed line. (b)
Momentum distribution (2) after $K=15$,
$N=10^{4}$ (solid line), $N = 5\times 10^{4}$ (crosses).
(c) The longitudinal (dotted line: $K=0$; shaded area: $K=15$) 
and (d) the transverse (dotted line: $K=0$; squares: 7; plusses: 15) 
spatial distributions for $N=10^{4}$ (thinner distributions) and 
$N = 5\times 10^{4}$ (broader distributions). The
transverse dynamics is frozen due to the small effective
nonlinearity. 
}
\label{fig:1}
\end{figure}

\begin{figure}
\centerline{\epsfig{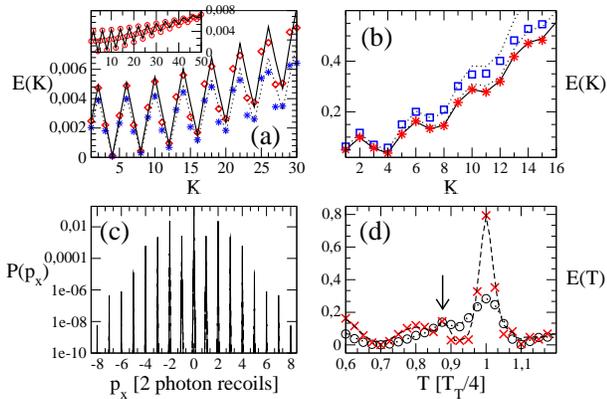}} 
\caption{(color online). Kinetic energies at $T=T_T/4$, same 
trap and pulse shape as in Fig. 1:
(a) 1D, $N=10^4$, $k=0.1$ (red diamonds) and $k=0.09$ (blue stars), and
(b) $k=0.45$ ($V_0/E_{\mb{R}}\simeq 42$), $N=10^{4}$ (1D, solid line;
3D red stars), 
and 3D  $k=0.48$ ($V_0/E_{\mb{R}}\simeq 45$), $N=10^{3}$ (squares).
(c) Momentum distribution (2) 
for 3D with $k=0.45$ after $K=15$. (d) A scan of the kinetic energy
vs $T$ for the 1D case from (b), after $K=10$ (circles) and 
$K=20$ (red crosses), with the corresponding data for the linear KR (dotted and
dashed lines).  The resonance at $T=T_T/4$ manifests itself
clearly, and a tiny peak of another 
higher-order resonance is marked by the arrow. The inset in (a)
confirms the correspondence between our 
method (circles) and a 1D fast Fourier transform 
evolution of the linear KR (solid line)
[a wave packet (circles) or an incoherent ensemble of plane waves (solid line)
was evolved respectively, with initial Gaussian momentum distribution 
-- $\sigma_{p_x}=0.026 \ \hbar k_L$ -- $k=0.09$]. 
For comparison, the linear KR ($p_{\mbox{\tiny init}}=0$)
in (a) for $k=0.1$ (solid line), $k=0.09$ (dotted line), and 
(b) $k=0.48$, $k=0.45$ (dotted line).
}
\label{fig:2}
\end{figure}

Figure 1 presents our results for the fundamental QR
of the KR at the Talbot time for Rb atoms,
$T=T_T= \pi \hbar /(2E_{\mb{R}}) 
\simeq 66.26\;\mu$s, with $k_L \simeq 8.1\times10^6 \rm \; m^{-1}$. 
Shown are the kinetic energy and the momentum
distribution of the condensate along the longitudinal direction. 
The energy is computed from the
momentum distribution, integrated over the transverse
directions, i.e.,
\begin{eqnarray}
E(K)&=& \frac12 \int dp_x p_x^2 P(p_x,K) \nonumber \\
\mbox{with}~~ P(p_x,K) &\equiv &\int dp_y dp_z 
|\psi(\vec{p},t=KT)|^2\;,
\label{eq:KE}
\end{eqnarray}
where $K$ denotes the number of kicks.
We present also the integrated spatial distribution
along the transverse direction [in $y$, or equivalently $z$, because of the
radial symmetry in Eq.~(1)]: $P_y(K)= \int dx dz |\psi(\vec{r},t=KT)|^2$.
As an example for higher-order QRs, which up to now have
never been resolvable experimentally, Fig. 2 shows data for the 
resonance at $T=T_T/4$. For all the parameters studied, the ballistic
quantum transport, with a quadratic growth $E(K) \propto K^2$, 
is clearly visible.  

Deviations from the idealized ballistic motion arise because of the
contributions of the trap potential and the nonlinearity. At long times,
the expanding condensate feels the harmonic trap potential,
and further acceleration is hindered by the trap. 
This effect is negligible for vanishing $\omega_x$.
Equating the longitudinal trap potential and the kinetic energy, and using
$p_{x} \lesssim \pi kK \hbar k_L$ at $T=T_T$ \cite{Izr,WGF}, we
estimate the kick number above which the trap dominates the dynamics
as $K \sim 2k_L \sqrt{\hbar/M\omega_x} \simeq 55$ for the parameters
of Fig.1.
More crucial is the small but finite initial condensate momentum spread 
($\sigma_{p_x} \ll 2\hbar k_L$ \cite{Phillips1999,Wilson2004AR}), which, 
after a characteristic time $t^*\propto 1/\sigma_{p_x}$, leads to a linear
increase of the energy $E(K) \propto K$ \cite{WGF}. This crossover sets
in at $K \simeq 10$ in Fig. 1 (a).
For small $\sigma_{p_x}$, the condensate ground state extends over
many lattice sites of the kick potential, which in turn makes a smaller
trap potential necessary for ideal ballistic motion. The interplay between
these situations is illustrated in Fig. 3, where we systematically vary the
number of atoms in the condensate. We stress that the effect of the
nonlinearity manifests itself 
indirectly via its influence on the initial state,
while the nonlinear interaction is negligible during the kick evolution. 
This finding is quite surprising, remembering that
the QRs correspond to exact phase revivals between kicks. On short time
scales, however, the perturbation induced by the nonlinearity cannot accumulate
a large enough dephasing, because even for the small kick strength 
$k<0.5$ it is at least 1 order of magnitude smaller than the kinetic
energy. Our results are consistent with a simplified 1D model
analysis of the QR, in the presence of a small nonlinear perturbation 
\cite{RWA2004}. This analysis, 
however, could not account for the exact nonlinear
wave packet evolution including the harmonic confinement. 

We finally note that for a large number of atoms 
$N\geq 5\times 10^4$ the results of our 1D model
and the full 3D computations differ [c.f. Fig. 3(a)]
because initially the condensate substantially
expands along the transverse directions. This is crucial for the
realization of ballistic quantum transport, since
for the same number of atoms in the condensate, the trap has less effect
in the 3D as compared with the 1D case, but, on the other hand, $\sigma_{p_x}$
in the 3D case is slightly larger. For the parameters of Fig. 3(a), the larger
$\sigma_{p_x}$ has a negligible influence, while the smaller initial spatial
extension allows the ballistic motion to survive
longer in the 3D than in the 1D case.

The nonlinearity manifests itself if we scan the kick period
over the fundamental QR and plot the kinetic energy at fixed 
$K$. The resonance peak shows a slight asymmetry, which does not occur
in the linear KR. The asymmetry decreases when (i)
reducing the longitudinal confinement [solid line as compared with
circles in Fig. 4(a)], or (ii) evolving the initial
condensate state without the nonlinear term [pyramids in Fig. 4(a)]. 
Again the trap potential hides the influence of the nonlinearity, but 
the observed asymmetry originates from both perturbations of the usual KR.
Any such perturbation is expected to introduce
an asymmetric peak shape, following a semiclassical analysis
specifically developed for the description of decoherence by spontaneous
emission \cite{WGF}. 

\begin{figure}
\centerline{\epsfig{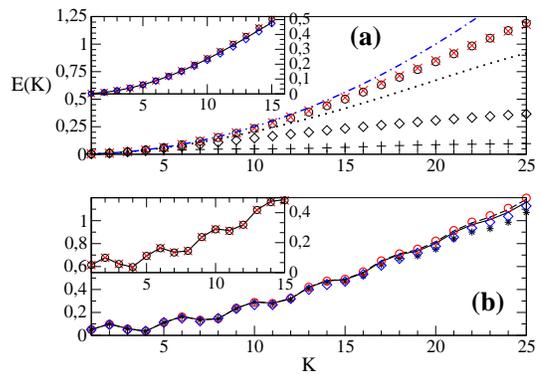}} 
\caption{(color online). Scan of the number of atoms (trap
as in Fig.1 if not stated otherwise): (a) for
$T=T_T$, and 1D  $N=5\times 10^3$ (crosses), $N=10^4$ (circles) 
$N=5\times 10^4$ (diamonds), $N=10^5$ (plusses); 
3D in the inset for $N=10^3$ (solid line),  $N=5\times 10^3$ (crosses),
$N=10^4$ (circles) $N=5\times 10^4$ (diamonds). The dot-dashed line  
shows the analytical result $E(K)=k^2K^2/4$ [linear KR, 
$p_{\mbox{\tiny init}}=0$], and the dotted line
1D data for $\omega_x/2\pi=5\rm \;Hz$, and $N=5\times 10^4$.
(b) $T=T_T/4$, $k=0.45$, 
1D for $N=10^2$ (stars), $N=10^3$ (solid line), 
$N=3\times 10^3$ (dashed), $N=10^4$ (circles), 
$N=5\times 10^4$ (diamonds); 
3D in the inset for $N=10^3$ (solid line), 
$N=5\times 10^3$ (crosses), $N=10^4$ (circles).
}
\label{fig:3}
\end{figure}

\begin{figure}
\centerline{\epsfig{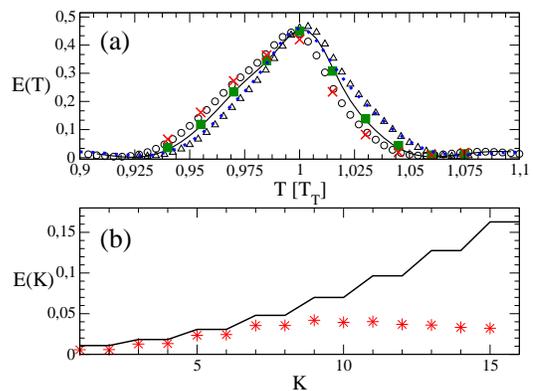}} 
\caption{(color online).
(a) Scan of the kinetic energy 
vs $T$ around $T=T_T$, parameters
as in Fig. 1, in comparison with
the linear KR (blue dots; $\sigma_{p_x} = 0.01 \ \hbar k_L$), after $K=14$;
1D ($N=10^4$: circles; $\omega_x /2\pi = 5\ \rm Hz$: solid line;
$N=0$ in kick evolution: pyramids),
and 3D ($N=10^4$: green squares; $N=5\times 10^4$: red crosses).
(b) Kinetic energy of 3D KHO with 
$T=250\ \mu$s, $\omega_x/2\pi=1 \rm \;kHz$, 
$\omega_{\mbox{\tiny r}}/2\pi = 5 \rm \;kHz$, and $k=0.1$ for $N=0$ 
(solid line) and $N=100$ (stars).
}
\label{fig:4}
\end{figure}

We can also tune our system described by the GPE (1) to the limit, in which 
the nonlinear interaction term becomes very important and even dominant
for small kick strength. Experimentally such a
situation can be realized either close to a Feshbach resonance (where the
scattering length for two-body atomic collisions can increase by orders of
magnitude \cite{PS2002}) or by simply increasing the strength of the harmonic
confinement. The latter situation has been analysed as a nonlinear
generalization of the KHO in \cite{Zoller2000,AR2002}. The analysis showed
clear signatures of the nonlinearity in the resonant transport regime of the 
KHO, i.e., for $T\omega_x=2\pi/r$ with $r=3,4,6$, where the quantum transport
is enhanced with respect to the classical one, much in the same way as at
the QRs of the KR \cite{QKHO}. 
In the KR studied above, the total energy -- corresponding to the
chemical potential in the stationary case \cite{PS2002} -- 
is almost entirely given by the kinetic energy obtained
from the kicks. In the KHO, the energy
distribution is very different: here the kinetic, the potential, and
the self energy (given by the nonlinear term) are of the same order
of magnitude, which leads to a nonlinearity-induced redistribution of energy 
also to the transverse degrees or freedom. 

Our numerical technique can be used to generalize the previous studies
of the nonlinear KHO to the full 3D GPE, 
properly including the coupling of the transverse dimensions which is
necessary for comparison with real-life experiments. Results on the KHO are
presented in Fig. 4(b), where we observe the destruction of the 
resonant motion at $T\omega_x=\pi/2$ 
already for a small number of atoms $N=100$, and 
after just a few kicks. In contrast to what was done in 
\cite{Zoller2000,AR2002}, the initial state need not be translated away from
the classically stable origin in phase space. It suffices to use the
relaxed ground state of the condensate to observe the impact of the now much
stronger effective nonlinearity due to the strong harmonic confinement. 

We have tested our data in different ways, making sure that our 
numerical codes produce stable results. 
In the 1D case and for $g_{1D}=0$, we compared them either with
analytical results for the linear KR \cite{Izr}, 
or with the much simpler evolution on a discrete grid, using 
a standard fast Fourier transform (FFT)  \cite{WGF,RWA2004}. 
The analytic growth rate of the kinetic energy for $p_{\mbox{\tiny init}}=0$ 
at $T=T_T$ is shown in Fig. 1(a). For 
$T=T_T/4$, in the inset of Fig. 2(a), we compare both evolutions 
for up to $K=50$ kicks. The FFT code does not take into account (i) the trap, 
(ii) the finite duration of the pulses, 
and (iii) the coherent evolution of the wave packet,
which explains the tiny deviations for 
$K\geq40$. The damping of the oscillations
is due to the nonzero $\sigma_{p_x}$, an effect which was experimentally
observed also for the anti-resonance $T=T_T/2$ in \cite{Wilson2004AR}, 
where the motion of the linear KR is perfectly periodic only for 
$p_{\mbox{\tiny init}}=0$.

In summary, we presented the first (3+1) dimensional treatment
of a Bose condensate driven by an external temporally periodic force, which 
controls the dynamics of the system. Ballistic motion is shown to be 
realizable over a substantial number of kicks, even in the presence of 
a weak harmonic confinement. Within the
framework of the 3D GPE, we have taken a first step
towards the study of higher-dimensional chaos induced by the nonlinear
coupling of the spatial dimensions, and future work will concentrate
on situations where the transverse degrees of freedom significantly contribute
to the dynamics.

This work was supported by MIUR, COFIN-2004, 
the Humboldt Foundation (Feodor-Lynen Program),
the Scuola di Dottorato G. Galilei, and the ESF (CATS).


\begin{thebibliography}{20}

\bibitem{Phillips1999}
L. Deng {\em et al.}, Phys. Rev. Lett. {\bf 83}, 5407 (1999).

\bibitem{Wilson2004AR}
G.J. Duffy, A.S. Mellish, K.J. Challis, and A.C. Wilson, 
Phys. Rev. A {\bf 70}, 041602(R) (2004).

\bibitem{CC}
S. P\"otting, M. Cramer, and P. Meystre, Phys. Rev. A {\bf 64}, 
063613 (2001);
J. Gong and P. Brumer, Phys. Rev. Lett. {\bf 88}, 203001 (2002).  

\bibitem{QRW}
See, e.g., D. K. W\'ojcik and J. R. Dorfman,  Phys. Rev. Lett. {\bf 90}, 
230602 (2003);
T. A. Brun, H. A. Carteret, and A. Ambainis, {\em idid.} {\bf 91}, 
130602 (2003).

\bibitem{expold}
M.G. Raizen, Adv. At. Mol. Opt. Phys. {\bf 41}, 43 (1999).

\bibitem{DT}
D.A. Steck, W.H. Oskay, and M.G. Raizen, Science 293, 274 (2001);
W.K. Hensinger {\em et al.}, Nature 412, 52 (2001).

\bibitem{AC} 
M.K. Oberthaler {\em et al.}, Phys. Rev. Lett. {\bf 83}, 4447 (1999).

\bibitem{Izr} 
F.M. Izrailev, Phys. Rep. {\bf 196}, 299 (1990).

\bibitem{QRexp} 
W.H. Oskay {\em et al.}, Opt. Comm. {\bf 179}, 137 (2000);
M.B. d'Arcy {\em et al.}, Phys. Rev. E {\bf 69}, 027201 (2004);
G. Duffy {\em et al.}, {\em ibid.} {\bf 70}, 056206 (2004);
S. Wimberger {\em et al.},  Phys. Rev. A (to be published),
e-print physics/0502061.
  
\bibitem{PS2002}
C.J. Pethick and H. Smith, {\it Bose-Einstein Condensation in Dilute Gases},
(Cambridge University Press, Cambridge, 2002); L. Pitaevskii and S. Stringari,
{\it Bose-Einstein Condensation}, (Oxford University Press, Oxford, 2003).

\bibitem{Zoller2000}
S.A. Gardiner {\em et al.}, Phys.~Rev.~A {\bf 62}, 023612 (2000).

\bibitem{ZLRN2004}
C. Zhang, J. Liu, M.G. Raizen, and Q. Niu, 
Phys. Rev. Lett. {\bf 92}, 054101 (2004);
B. Mieck and R. Graham, J. Phys. A {\bf 37}, L581 (2004).

\bibitem{QKHO}
G.M. Zaslavsky {\em et al.}, {\em Weak chaos and quasi-regular patterns}
(Cambridge University Press, 1992); F. Borgonovi and L. Rebuzzini, 
Phys. Rev. E {\bf 52}, 2302 (1995);
A.R.R. Carvalho and A. Buchleitner, Phys. Rev. Lett. {\bf 93}, 204101 (2004).

\bibitem{Mannella1998}
E. Cerboneschi {\em et al.}, Phys. Lett. A {\bf 249}, 495 (1998).

\bibitem{O1998}
M. Olshanii, Phys. Rev. Lett. {\bf 81}, 938 (1998).

\bibitem{Ess2003}
H. Moritz, T. Stoferle, M. K\"ohl, and T. Esslinger, Phys. Rev. Lett.
{\bf 91}, 250402 (2003).

\bibitem{WGF} 
S. Wimberger, I. Guarneri, and S. Fishman,
Nonlinearity {\bf 16}, 1381 (2003); Phys. Rev. Lett. {\bf 92}, 084102 (2004).

\bibitem{RWA2004}
L. Rebuzzini, S. Wimberger, and R. Artuso, 
Phys. Rev.~E {\bf 71}, 036220 (2005). 

\bibitem{AR2002}
R. Artuso and L. Rebuzzini, Phys. Rev. E {\bf 66}, 017203 (2002).

\end{thebibliography}
\end{document}